# Ruddlesden-Popper defects act as a free surface: role in formation and photophysical properties of $CsPbI_3$

Weilun Li[1], Qimu Yuan[2], Yinan Chen[2], Joshua R. S. Lilly[2], Marina R. Filip[2], Laura M. Herz[2,3], Michael B. Johnston[2], Joanne Etheridge[1,4,5*]

[1] School of Physics and Astronomy, Monash University, VIC, 3800, Australia

[2] Department of Physics, University of Oxford, Clarendon Laboratory, Parks Road, Oxford OX1 3PU, United Kingdom

[3] Institute for Advanced Study, Technical University of Munich, Lichtenbergstrasse 2a, D-85748 Garching, Germany

[4] Monash Centre for Electron Microscopy, Monash University, VIC, 3800, Australia

[5] Department of Materials Science and Engineering, Monash University, VIC, 3800, Australia

* Joanne Etheridge

**Email:** joanne.etheridge@monash.edu

**Author Contributions:** M.B.J. and J.E. designed the study and supervised the project. W.L. designed and carried out TEM experiments and analysed data. Q.Y. optimised and prepared specimens and performed XRD, PL, TRPL measurements. Y.C. and M.R.F. performed DFT calculations. J.R.S.L. and L.M.H. performed in-situ TRPL/TRMC measurements and associated modelling. W.L. and J.E. prepared the manuscript. All authors contributed to the discussion of the results and revision of the manuscript.

**Competing Interest Statement:** The authors declare that they have no competing interests.

## Abstract

The perovskite semiconductor, $CsPbI_3$, holds excellent promise for solar cell applications due to its suitable bandgap. However, achieving phase-stable $CsPbI_3$ solar cells with high power conversion efficiency remains a major challenge. Ruddlesden-Popper (RP) defects have been identified in a range of perovskite semiconductors, including $CsPbI_3$. However, there is limited understanding as to why they form or their impact on stability and photophysical properties. Here we increase the prevalence of RP defects with increased Cs-excess in vapour-deposited $CsPbI_3$ thin films and observe superior structural stability but inferior photophysical properties. Significantly, using electron microscopy, we find that the atomic positions at the planar defect are comparable to those of a free surface, revealing their role in phase stabilisation. Density functional theory (DFT) calculations reveal the RP planes are electronically benign, however, antisites observed at RP turning points are likely to be malign. We therefore propose that increasing RP planes while reducing RP turning points could offer a breakthrough for improving both phase stability and photophysical performance. The formation mechanism revealed here may well apply more generally to RP structures in other perovskite systems.

## Main Text

## Introduction

In the past two decades, metal halide perovskite semiconductors have generated considerable attention for their potential application in solar cells [1–4]. Presently, the majority of high-efficiency perovskite solar cells utilise organic cations, primarily methylammonium ($MA^+$) and formamidinium ($FA^+$). However, the volatility of these organic molecules promotes material degradation and requires complex solvent-engineering, presenting a significant obstacle to the long-term viability of organic-inorganic hybrid perovskite solar cells [5,6]. In recent years, all-inorganic $CsPbI_3$ perovskite solar cells have emerged as a subject of intense interest owing to their suitable band gap and enhanced environmental stability relative to the organic-inorganic hybrid systems [7–9]. However, the perovskite phases (α, β, and γ) are metastable and quickly transforms into a thermodynamically favoured non-perovskite δ phase, which is not photoactive and hence is not suitable for solar cells [10]. While various strategies have been explored to prepare $CsPbI_3$ solar cells with both high efficiency and structural stability, such as composition engineering, interfacial passivation, and template methods [11,12], it remains a major challenge. In the meantime, vapour deposition is a technique well-suited for the conformal and homogenous coating of metal halide perovskites, whilst exhibiting excellent compatibility with existing industrial production lines for commercial upscaling [13,14]. This solvent-free method has also allowed for the direct formation of meta-stable γ-phase $CsPbI_3$ with a low processing-temperature [15–20].

Ruddlesden-Popper (RP) phases are derivatives of the perovskite structure, possessing n-layers of $ABX_3$ interleaved with, typically, AX rock-salt layers to give a general formula of $A_{n+1}B_nX_{3n+1}$ [21,22]. RP phases have been reported across a wide variety of halide perovskites [23–25]. Some are intentionally produced by incorporating large cations, including molecules, as spacers [26], while others arise as defects within the perovskite structure, often known as RP defects [25,27,28]. Despite numerous reports, the detailed atomic-scale structure of RP defects and their role in the stability and optoelectronic properties of perovskite solar cells are not yet clear.

In this study, we control the prevalence of RP defects by manipulating the nominal ratios between CsI and $PbI_2$ in vapour co-deposited $CsPbI_3$ films and determine their effect on the corresponding phase-stability and optoelectronic properties. We observe that an excess of Cs is necessary for stabilising $CsPbI_3$ in its photoactive γ perovskite phase and that Cs-rich compositions promote the formation of RP planar defects which have a locally Cs-rich structure.

We examine the impact on structure by using quantitative transmission electron microscopy to measure the atomic positions and associated strain at the RP plane. Remarkably, we find the RP plane is comparable in structure to a free surface. This, together with measurements of a related 90° rotation boundary, allows us to identify the mechanism that drives the formation of RP defects and thereby understand how they achieve structural phase-stability in $CsPbI_3$.

We also examine the impact on photophysical properties by measuring the prevalence of RP defects and correlating with the measured photophysical properties. We use our measurements of the atomic structure and composition of the RP planes and RP turning points to understand their respective impact on electronic structure. We compare density functional theory (DFT) calculations of the measured RP planar defect structure with the defect-free structure. Using these structural measurements and calculations, we reveal the critical and different role of RP planes versus RP turning points in photophysical properties.

Together, these insights open a path for stabilising the γ- $CsPbI_3$ perovskite phase while optimising photophysical performance via RP defect engineering.

## Results

### Crystallographic phase and microstructure versus Cs:Pb ratio

We first characterised the microstructure of dual-source vapour-deposited $CsPbI_3$ films using X-ray diffraction (XRD) and low-dose, four-dimensional scanning transmission electron microscopy (4D-STEM) (with methodology and details given in **Supplementary Note 1** and **2**). The nominal Cs:Pb ratio was varied from 0.85 to 1.25, denoted as Cs-0.85, Cs-1.0, Cs-1.1 and Cs-1.25 for brevity.

The XRD data was obtained immediately after fabrication and is consistent with the formation of $CsPbI_3$ in an orthorhombic γ phase (space group: Pbnm) across all compositions, as shown in **Fig. S2**.

The thin TEM films were examined ~5 days after fabrication (stored until then in dry nitrogen and darkness). Using 4D-STEM, the γ-$CsPbI_3$ phase was detected only in the Cs-rich thin TEM films, with Pb-rich films Cs-0.85 and stoichiometric Cs-1.0 displaying a δ-$CsPbI_3$ phase (space group: Pmnb), as shown in **Fig. 1** and **Fig. S3**. This γ-to-δ phase transition is expected over time due to the known instability of the γ-phase for stoichiometric or Pb-rich $CsPbI_3$. The improved stability with Cs-excess was also observed in the thermal stability (85°C, $N_2$ atmosphere, dark) measurements of as-prepared γ-$CsPbI_3$ thin films, which show extremely poor stability for stoichiometric Cs-1.0 and significantly improved stability for Cs-1.25 (as detailed in **Supplementary Note 3**).

The XRD and TEM results also show a transition from a random grain orientation to a strongly preferred out-of-plane $[1\bar{1}0]$ texture with increasing Cs excess, as well as improved uniformity in grain size and shape (**Supplementary Note 2**).

To investigate further the phase and atomic structure of $CsPbI_3$ films, safe dose ($5.7 \times 10^3$ e/A$^2$), atomic-resolution STEM-ADF experiments were performed (see **Supplementary Note 4** for a discussion of safe electron dose for RP defects). STEM-ADF images in **Fig. 1(A-B)** confirm the Cs-0.85 and Cs-1.0 TEM films have the non-perovskite δ-$CsPbI_3$ phase**.**

The orthorhombic perovskite γ-$CsPbI_3$ phase is observed in the Cs-rich TEM specimens in **Fig. 1D**, Cs-1.1 and Cs-1.25, consistent with the XRD data. In addition, planar defects parallel to both the (002) and (110) planes are common. As we will demonstrate later, these planar defects have an atomic structure consistent with the RP phase [29]. These defects often include steps and 90° turns and, in some instances, form a loop inside the grain. For clarity, hereon, we will call the planar component of the defect, “RP planes”, the turning points “RP TPs” and one- or -two- unit cell steps, “RP steps”.

It is important to note that halide perovskites are extremely sensitive to electron irradiation [6,30–32]. Comprehensive investigations of structure integrity of RP defects have been conducted and discussed in **Supplementary Note 4**. The electron dose used throughout this study is well below the beam damage threshold to be sure that the defect and its atomic configuration is intrinsic to the material and has not been created or modified by the electron beam.

**Atomic structure of RP planes: 90° boundary, Cs-displacements and octahedral tilt relaxation.**

Two types of RP planar defect are observed, which we will call Type-90 and Type-0 (relating the presence or absence of a 90° rotation boundary, respectively). Both defects can be interpreted as the insertion of an additional Cs plane or the removal of a Pb plane, resulting in a locally Cs-rich composition. We measure their detailed atomic configuration here, with the key features summarised in **Box 1.**

The structure of the Type-90 defect is examined first in **Fig. 2.** It is evident that the defect creates a large gap between two distinct crystal domains, one shifted by half a unit cell relative to the other in the direction parallel to the defect line, resulting in a transition from Pb/I columns to Cs columns across the defect. Furthermore, in approximately 20% of defects, there is an in-plane switch of crystal axis from [001] (left domain) to [110] (right domain) (identified from the Fourier transform of the STEM-ADF images in **Supplementary Note 5**). Thus, the Type-90 RP plane acts as a 90° domain boundary, as illustrated in **Fig. 2G**. Remarkably, given the switch of in-plane axes across the RP gap, the crystallographic plane is necessarily different on each side of the gap, namely, (001) on one side and (110) on the other (as shown in **Fig. 2G**).

We measured the atomic positions adjacent to the Type-90 RP plane (from **Fig. 2C** and electron scattering calculations in **Supplementary Note 6)**. We find that the Cs columns closest to the RP plane are displaced significantly towards the interior of the domain, in the direction perpendicular to the RP plane, reducing the Cs-Cs column distance relative to the bulk. **Fig. 2D** illustrates the displacement vectors of Cs columns in the vicinity of the RP plane. In the left-hand domain with the (001) boundary, the Cs columns are displaced by the same amount, so the entire first (001) layer of Cs atoms adjacent to the gap moves collectively inward towards the interior. In the righthand domain with the (110) boundary, the magnitude of the displacement of the Cs columns adjacent to the gap *alternates.* Notably, this results in the first (110) layer of Cs atoms adjacent to the gap being located on exactly the *same* (110) plane (red line in **Fig. 2G**), *unlike* the Cs columns within the domain interior which lie on alternate (110) planes positioned either side of the corner of the octahedra, as shown by the dotted and dashed red lines in **Fig. 2G**.

Given the observed Cs displacements on each side of the RP gap, we now consider whether there is any accompanying change in the octahedral tilting near the gap. The orthorhombic structure of γ-$CsPbI_3$ entails octahedral tilts in all three axes, as illustrated in **Fig. 2E**. The ellipticity map shown in **Fig. 2F** reveals there is a relaxation of the octahedral tilt angle each side of the RP gap, as seen from the reduced vector length/intensity in the schematic. The ellipticity map also confirms there is an in-plane switch of crystal axis between [001] and [110], consistent with the 90° crystal rotation across the gap, as described above.

Finally, we measured the exact magnitude of the gap, the Cs displacement and octahedral tilt, by quantifying STEM-ADF images against electron scattering calculations (**Supplementary Note 6**). We determined that the Cs-Cs column distance on the (100) RP plane reduces from 6.3 Å in the interior to 5.8 Å at the interface, while on the (110) RP plane the reduction alternates between 0.2 Å and 0.5 Å. Similarly, the octahedral tilt angle relative to the bulk was observed to reduce by 33%. The gap distance, as defined in **Fig. 2G**, is 3.5 Å.

We also examined the atomic structure of the Type-0 RP planar defect in **Supplementary Note 7**. In this type, there is no rotation of the domain each side of the RP gap. Otherwise, the Type-0 planar defect is essentially the same as the Type-90 RP planar defect, comprising contraction of the Cs-Cs column distance perpendicular to the defect plane, reduction of the octahedral tilt angle and a comparable gap size.

The key structural features of the RP planar defects determined above are summarised in **Box. 1.**

- An additional Cs-plane, giving a local composition of $Cs_2PbI_4$
- A gap measuring g = 3.5 Å (see **Fig. 2G**)
- A half-unit cell shift parallel to the gap
- Cs-displacements up to 0.5 Å towards the domain interiors
- Octahedral tilt relaxation in the first layer adjacent to the RP gap
- The first layer of Cs atoms lie 'flat' on the (100) or (110) planes each side of the RP gap
- In Type-90 RP, a 90°crystal rotation across the RP-gap with different planar interfaces

**Box. 1. Atomic structural features of RP planar defects measured and refined from TEM experiments and simulations.**

### Atomic structure of RP Turning Points: Antisites and dangling bonds

In addition to the RP planes, we frequently observed RP turning points. Careful analysis of STEM images reveals that antisite defects at the turning points are common. In the STEM-ADF image, **Fig. 3A**, the nominal position of the Cs columns at some RP turning points show a significantly higher intensity than for the Cs column positions in the bulk. Given the relative atomic number of Pb and Cs, these higher intensity maxima can be attributed to the presence of Pb in these columns, suggesting the formation of $Pb_{Cs}$ antisite defects.

We examined various manifestations of RP defects, including those originating from grain boundaries (**Fig. 3A**), forming RP loops (**Fig. 3B**), or associated with a 90°crystal rotation (**Fig. 3C**). In all of these cases, the presence of antisite defects at turning points was often observed.

### Relationship between prevalence of RP defects and Cs:Pb ratio

We estimated the prevalence of RP planes and RP turning points versus Cs:Pb ratio. To quantify the measurement of prevalence, we used the total area of RP planar defects and the total columnar length of RP turning points within the total volume examined (as detailed in **Supplementary Note 8**). Many fields of view (> 20 grains) were analysed to ensure good statistics. The resulting statistics indicate that the prevalence of both RP planes and RP TPs increases significantly with increasing Cs:Pb ratio (**Fig. 3D**). Given the RP plane is in effect an additional Cs-plane, giving a local composition of $Cs_2PbI_4$, it is expected that Cs-excess is associated with increased prevalence of the RP defects.

### Formation mechanisms of RP defects: the gap acts as a free surface

To provide insights into the mechanisms driving the formation of RP defects in Cs-rich $CsPbI_3$ films, we consider separately two key features of the RP defect structure: the gap (always present) and the 90° boundary (sometimes present). To do this, we measure two separate structures; a free surface in Cs-rich $CsPbI_3$ film and a 90° boundary *without* a gap nor a half-unit cell shift in the stoichiometric $CsPbBr_3$ (choosing the bromine homologue because of the phase instability of stoichiometric $CsPbI_3$).

The atomic structure of the $CsPbI_3$ film at a free surface, that is, at an interface with the vacuum, was examined in **Fig. 4A**, which shows a rare pinhole in the Cs-1.1 $CsPbI_3$ film. At this surface, Cs displacements and octahedral relaxation are evident, which we measured to be comparable to the RP planar defects, as illustrated in **Fig. 4(B-D)**. *Remarkably, this suggests that the gap at the RP plane acts as a free surface*. This is a critical observation with implications both for phase stabilisation and photophysical properties.

It has been calculated that there is significant internal strain inherent in orthorhombic γ-$CsPbI_3$ films [33]. The observation here that the RP gap acts as a free surface provides an underpinning reason as to why the RP defect stabilises the γ-$CsPbI_3$ phase in films, namely, by reducing this internal strain, which we discuss later. Moreover, the Cs excess facilitates the formation of the defect by enabling the additional Cs atomic plane, enhancing stability.

### RP gap relieves strain at grain boundaries: Example, the 90° (110)/(100) domain boundary

In the orthorhombic phase of γ-$CsPbI_3$ (and γ-$CsPbBr_3$), the planar distance of (110) planes is different to (001) planes. Consequently, significant lattice mismatch would be expected across a 90° (110)/(100) domain boundary, as illustrated schematically in **Fig. 4F**, inducing significant tensile and sheer strain. We hypothesise that in the presence of excess Cs, the Type-90 RP planar defect can form, with the gap acting as a free surface, removing this strain.

We explore this hypothesis by measuring the degree of tensile and sheer strain in the *absence* of a gap at a 90° (110)/(001) domain boundary in the stoichiometric, phase-stable homologue γ-$CsPbBr_3$ films (prepared in the same way as γ-$CsPbI_3$ films, as described in **Supplementary Note 9**). (Such observations are not possible in phase unstable stoichiometry γ-$CsPbI_3$ films.) This domain boundary is in the field of view in the STEM-ADF image in **Fig. 4E** but is easily missed without careful analysis (see **Supplementary Note 10**) and has not been reported previously, to the best of our knowledge.

Unlike the Type-90 RP planar defect in γ-$CsPbI_3$ films, there is no gap between the two domains and no half-unit cell shift. The lattice mismatch induces tensile and shear strain in the direction perpendicular and parallel to the boundary. The resultant strain is evident from measurements of the Pb-Pb distance as illustrated in **Fig. 4F**. **Fig. 4(H-J)** shows a gradual increase in Pb-Pb distance across the boundary in the direction perpendicular to the boundary, together with a sudden decrease at the boundary in the parallel direction. Significantly, there is also no Cs displacement at the domain boundary relative to the interior in **Fig. S14**, suggesting this only happens when a RP planar defect is formed, and a free surface is created at the gap.

The observations above are consistent with the hypothesis that the gap at the RP defect plays a significant role in the relief of internal strain associated with incoherent grain boundaries. The specific example of the (110)/(100) 90° domain boundary is a case in point. It forms in the transition from cubic to orthorhombic γ-$CsPbI_3$ when the lattice parameters along the [001] and [110] directions (i.e. lattice parameter b and c in its cubic phase) cease to be equivalent.  The tensile strain from the lattice mismatch provides a driving force for the two domains to be separated by a gap, while the shear strain provides a driving force for the half-unit cell shift parallel to the boundary.

**Photophysical properties versus Cs:Pb ratio**

We examined the optoelectronic properties of 35 nm thick vapour-deposited γ-$CsPbI_3$ films with different nominal Cs:Pb ratios on z-cut quartz substrates (**Fig. 5** and **Supplementary Note 11**). Similar measurements of device-thickness γ-$CsPbI_3$ films and for films deposited on TEM-grids are shown in **Fig. S17** and **Fig. S18**, which illustrate identical trends. As Cs excess is necessary for the stabilisation of γ-$CsPbI_3$, studies were performed on the near-stoichiometric Cs-1.03 films, together with the more phase-stable Cs-1.1 and Cs-1.25 films. Photoluminescence (PL) spectra in **Fig. 5A** and **Fig. S15** indicate no considerable shift in the emission peak wavelength with compositional variation. Notably, the unnormalized PL peak intensity reduces significantly with increasing Cs-excess. Similarly, effective lifetimes extracted from time-correlated single photon counting measurements (TCSPC) in **Fig. 5B** and **Fig. S16** exhibit a decreasing trend with rising Cs concentration. These observations suggest that the use of a more Cs-rich precursor stoichiometry leads to an increase in the rate of non-radiative charge-carrier recombination.

As illustrated in **Fig. 5C** and **Fig. S19**, we further probed charge-carrier recombination dynamics via the concurrent acquisition of time-resolved photoluminescence (TRPL) and time-resolved microwave conductivity (TRMC) transients measured under identical optical excitation conditions [34].  For Cs-1.1 and Cs-1.25 films, a significant remnant photoconductivity component (**Fig. 5C**), is observed with a concomitant elimination of PL signal (**Fig. S19**). Such a coupled response is indicative of trapping via species-specific retaining trap states: one charge-carrier species is preferentially trapped in a retaining trap state, with the other species thus being unable to recombine (the source of the remanent photoconductivity signal) [34].

To investigate quantitative differences in recombination behaviour between the three films studied, TRMC and TRPL transients were simultaneously fit with a dynamic recombination model (see Supplementary Notes 1 for details). As demonstrated in **Fig. 5C** and **Fig. S19**, the model was able to reproduce both the measured photoconductivity and photoluminescence response. **Fig. 5D** shows the total trap-mediated recombination rate extracted from the model fitting as a function of the nominal Cs:Pb ratio: a monotonic rise is observed in the rate of charge-carrier recombination via trap states as Cs-excess is increased.

Interestingly, there was no discernible change in effective electron-hole sum charge-carrier mobility, as determined from TRMC measurements, between the three nominal compositions considered (**Tab. S3**).

In summary, the above results indicate that increasing Cs excess correlates with an increase in the rate of trap-mediated recombination, suggesting a concomitant rise in the formation of defect states [35,36]. Given Cs excess encourages the formation of RP defects, it is pertinent to ask whether RP defects could be acting as trapping sites for photoexcited charge-carriers in γ-$CsPbI_3$? In the structural observations above, we observe that RP planes act as free surfaces and RP turning points often contain antisites. We thus consider the impact on electronic structure below using DFT calculations.

## Density Function Theory calculations of RP defects

We next performed DFT calculations (including spin-orbit coupling effects; see **Methods** for computational details) to predict the effects of RP planar defects and their local atomic structure on the electronic band structure. As illustrated in **Fig. 5E** and **Fig. S20**, four structural models were built, namely the "bulk model" (perovskite without RP defect); a "naïve model" (perovskite with an incorrect RP model whereby the atomic positions are unchanged from the bulk perovskite); a "Type-0 RP model" and a "Type-90 RP model", with atomic positions as measured using TEM above, incorporating Cs displacement, octahedral relaxation in both models and 90° crystal rotation in the Type-90 RP model.

To probe the reliability of our atomistic models, we calculate band gaps for different RP planar defect concentrations by varying the number of octahedral layers (slab thickness) from 3 to 8 for both the Type-90 RP and naïve RP models. In **Fig. 5F** and **Supplementary Note 12**, we show that as the slab thickness increases, the calculated band gap converges towards the band gap of the bulk, with the gap of the 8-layer defect model only 0.09 eV away from that of the orthorhombic bulk structure. This is consistent with the expectation that the effect of defect interfaces will be less pronounced with decreasing concentration. We note that all DFT calculated band gaps are expectedly underestimated with respect to experiment by more than 1 eV, as documented extensively in the literature [37,38]. Furthermore, we find that band edge shapes do not change significantly as the slab thickness changes. These tests confirm that our analysis of the electronic structure for the 4-layer thick structure discussed in the following is physically meaningful and relevant for the description of the optoelectronic properties of $CsPbI_3$ thin films inclusive of RP defects.

To investigate further how the defect interface affects the electronic properties, we calculate the band structures of the 4-layer Type-0 RP, Type-90 RP and naïve-RP models, as well as an 8 × 2 × 2 $CsPbI_3$ bulk supercell, shown in **Fig. 5G**.

In the band structure of the bulk $CsPbI_3$ supercell, we see more intricate band crossings along X to Γ than Γ to Y, which can be attributed to band folding. Comparing the band structure of the bulk supercell with that of the naïve-RP model, we observe that the band dispersion along X to Γ is completely suppressed, due to the incorporation of the RP defect (i.e. gap and lattice shift at the RP plane). Furthermore, the VBM (valence band maximum) and CBM (conduction band minimum) in the naïve-RP model is a four-fold degenerate state (including spin-degeneracy), by contrast with the band edges of the bulk structures which exhibit a two-fold degeneracy (due to spin only).

We next compare the three RP models. We find that the Cs displacement and octahedral relaxation (naïve RP vs Type-0 RP) lead to a marginal reduction in the band gap by less than 50 meV. The 90°crystal rotation across the gap breaks the four-fold degeneracy of the VBM and CBM into two sets of bands separated by approximately 0.2 eV (Type-0 RP vs Type-90 RP), due to a change in the electrostatic potential upon the 90° rotation of the crystal. The band gap reduction (as depicted in **Fig. 5G**) and the breaking of the 4-fold degeneracy by the 90° crystal rotation is consistently observed across all RP defect interface systems with different slab thickness, confirming that the origin of this effect is due to the subtle structural changes at the RP interface.

Overall, the band structures for the three RP models shown in **Fig. 5G** exhibit very similar band edge shapes and do not appear to introduce any trap states in the band gap, consistent with previous studies [27,28,39]. Therefore, we conclude that these DFT results show no evidence that the RP planar defects and their associated structural variations would have adverse effects on the optoelectronic properties of $CsPbI_3$ thin films.

## Discussion

Measuring and understanding the atomic structure of RP defects enables us to elucidate their role in the phase stability and photophysical properties of $CsPbI_3$ with varying Cs:Pb ratios.

We first consider the impact on phase stability. TEM results reveal that RP planar defects possess an additional plane of Cs and I ions, resulting in a $Cs_2PbI_4$ local composition. This is consistent with the observations of RP defects in Cs-rich conditions and the increase in the prevalence of RP defects with increasing Cs excess [40,41]. In other words, the introduction of excess Cs promotes the formation of RP defects to accommodate the CsI-rich non-stoichiometric composition.

To understand how RP defects might impact phase stability, we consider how the RP planar defect serves to reduce internal strain. During the preparation and crystallisation of polycrystalline $CsPbI_3$ films, rapid crystal growth at relatively low temperature promotes the formation of defects such as grain boundaries and other intragrain defects. Due to the small size of Cs ions, $CsPbI_3$ goes through a phase transition by tilting the $[PbI_6]^-$ octahedra from α(cubic) to β(tetragonal) and eventually γ(orthorhombic) at room temperature. It has been shown that the spontaneous strain introduced by the lattice distortion during the phase transition is most pronounced in γ-$CsPbI_3$ compared with other halide perovskites, such as $CsPbBr_3$, $CsPbCl_3$, $MAPbI_3$ and $FAPbI_3$ [33]. Strain originated from both incoherent boundaries and octahedral distortions, contributes to the large internal strain and inherent phase instability of γ-$CsPbI_3$. This corresponds to cases of stoichiometric $CsPbI_3$ and Pb-rich $CsPbI_3$, where the as-prepared γ-$CsPbI_3$ rapidly transforms into undesired δ-$CsPbI_3$.

While the internal strains associated with the phase-stable bromine homologue, orthorhombic γ-$CsPbBr_3$, are calculated to be less than for phase-*un*stable stoichiometric γ-$CsPbI_3$ [33], our TEM measurements show there is still significant strain associated with the (001)/(110) 90º domain boundaries in γ-$CsPbBr_3$. Such boundaries arise in the tetragonal to orthorhombic phase transition when lattice parameters along the [001] and [110] directions (i.e. lattice parameter b and c in its cubic phase) cease to be equivalent. Remarkably, in Cs-rich γ-$CsPbI_3$, these domain boundaries are accommodated with no measurable strain as they only exist in association with the RP defect where we have found its gap acts as a free surface.

This boundary provides a specific example of how the formation of RP planar defects separates otherwise strained crystals into sub-domains, relieving the strain with the introduction of the RP gap which acts like a free surface. RP planar defects typically penetrate through the crystal together with RP turning points and steps, effectively resolving the lattice mismatch between domains in all dimensions through the provision of a free surface. Consequently, Cs-excess in γ-$CsPbI_3$ suppresses lattice strain and phase instability by promoting RP defect formation.

We have observed a strong correlation between the prevalence of RP defects and an increase in the rate of non-radiative recombination. To understand whether this is a coincidence or causative, we consider the structural observations and DFT calculations above. The RP planar defect acts as a free surface with minimal related strain. Furthermore, DFT calculations based on the real structure of the RP planar defect suggests that its impact on the electronic structure, while measurable, is likely to be relatively benign with respect to photophysical properties. This is supported by our observation that the effective electron-hole sum charge-carrier mobility, determined from TRMC measurements, exhibits no discernible change as the nominal Cs:Pb ratio (and hence RP defect density) is altered.

On the other hand, at the RP turning points we observed many $Pb_{Cs}$ antisite defects. Pb atoms located at these antisites (i.e., original Cs sites) are likely to have dangling bonds due to the significantly larger distance to halide atoms on the opposite side of the RP defect. Recent DFT calculations have suggested that the formation of $Pb_{Cs}$ antisite defects [42] or the presence of dangling Pb bonds associated with halide vacancies could result in deep-level trap states [39] that are likely to be detrimental to photophysical properties. This aligns well with our observation that increasing the prevalence of RP turning points, and hence antisites and dangling bonds, corresponds with a decrease in optoelectronic performance, as shown in **Fig. 5**.

## Conclusions

The formation of RP planar defects requires an additional Cs plane at the defect interface and is hence promoted with excess Cs. The RP planar defect provides a mechanism to reduce internal strain and stabilise the orthorhombic γ-$CsPbI_3$ phase through the provision of a free internal surface via the RP gap. This mechanism may well apply to the many other perovskite systems that exhibit RP defects, in ferroelectrics, superconductors and batteries [43–46]. DFT calculations suggest the RP planar defect is likely to be electronically benign with respect to photophysical properties. However, RP turning points, which also increase with Cs-excess, support antisite defects with Pb dangling bonds lead to deep trap states detrimental to photophysical properties. Collectively, these observations suggest that tuning Cs-excess and growth parameters to deliberately incorporate RP planar defects without RP turning points in orthorhombic γ-$CsPbI_3$ may ensure phase stability without impacting optoelectronic properties, opening a new avenue for the development of high-performance solar cells with phase-stable $CsPbI_3$.

## Acknowledgments

Financial support from the Australian Research Council (ARC) is appreciated. J.E. acknowledges ARC Discovery Project DP200103070 and ARC Laureate Fellowship FL220100202. The authors acknowledge use of facilities within the Monash Centre for Electron Microscopy, a node of Microscopy Australia. The Thermo Fisher Scientific Spectra φ TEM was funded by ARC LE170100118 and the FEI Titan$^3$ 80-300 FEG-TEM was funded by ARC LE0454166. Q.Y. acknowledges the support of Rank Prize through a Return to Research grant. J.R.S.L. and L.M.H. acknowledge financial support from the Engineering and Physical Sciences Research Council (EPSRC). J.R.S.L. thanks Oxford Photovoltaics Ltd. for additional support as part of an EPSRC Industrial CASE studentship.

## Figures and Tables

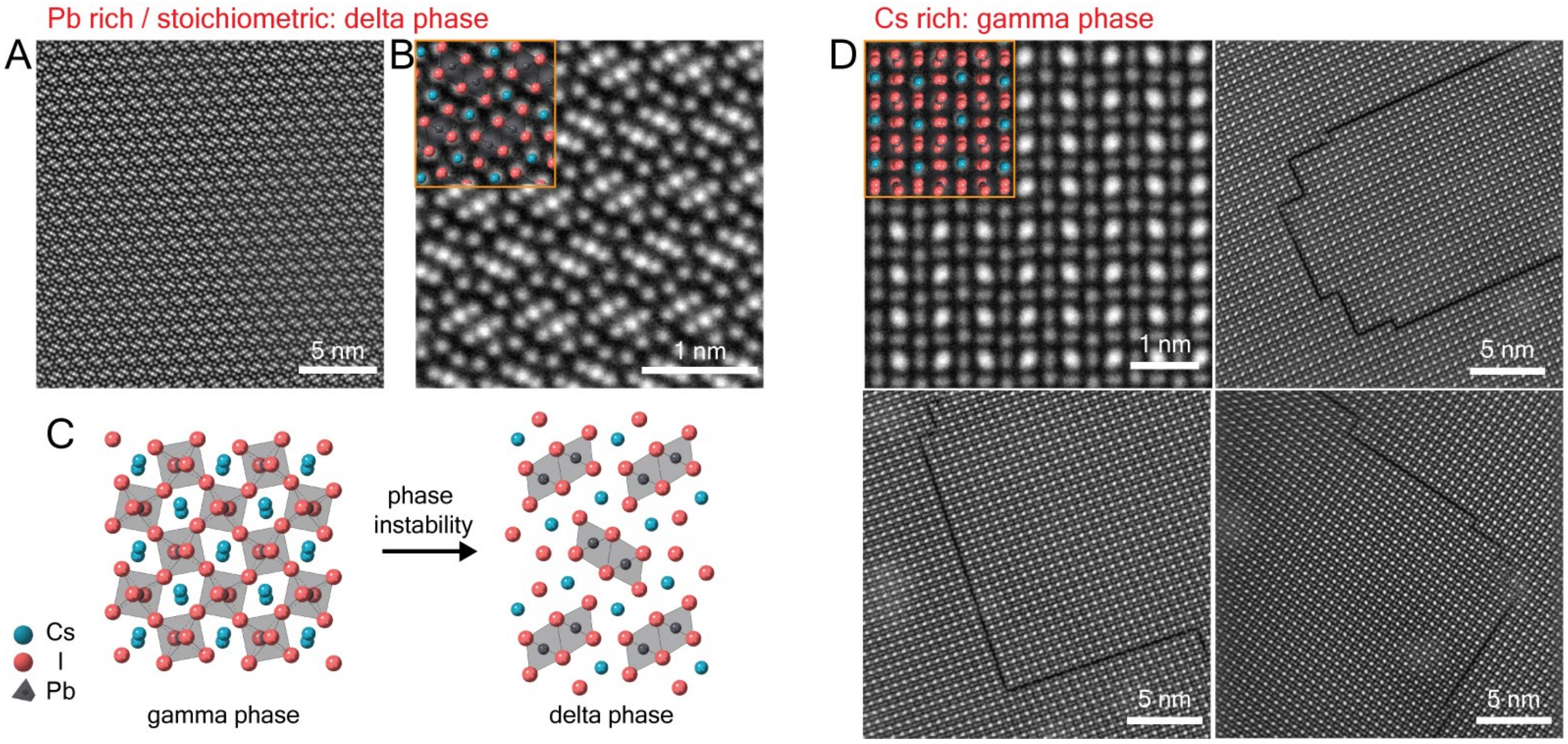

**Fig. 1. Atomic resolution TEM images of $CsPbI_3$ films with different Cs compositions. (A, B)** STEM-ADF images of the δ-$CsPbI_3$ phase in Cs-0.85 films in the [100] zone axis. **(C)** A schematic diagram of the phase transition from the initial γ-$CsPbI_3$ into δ-$CsPbI_3$. γ-$CsPbI_3$ is viewed in the [001] zone axis (Pbnm). δ-$CsPbI_3$ is viewed in the [100] zone axis (Pmnb). **(D)** STEM-ADF images of the γ-$CsPbI_3$ (top left) and RP defects in Cs-rich specimens, Cs-1.1 and Cs-1.25, viewed in the $[1\bar{1}0]$ zone axis. Note, the highest intensity maxima in the atomic-number sensitive STEM-ADF images correspond to columns containing Pb/I atoms (alternating in the beam direction), while the lower and comparable intensity maxima correspond to pure I and pure Cs atom columns.

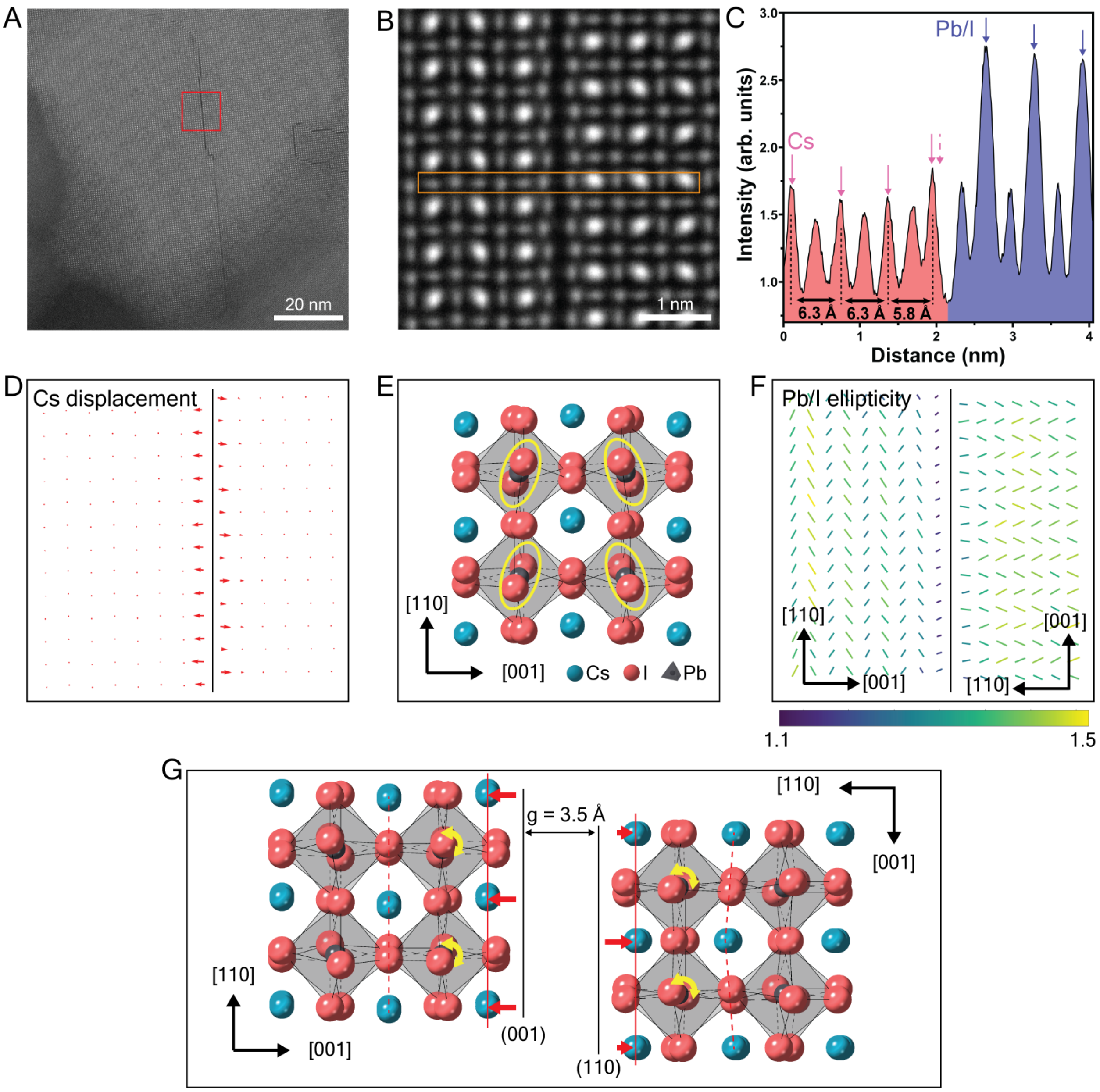

**Fig. 2. Quantitative analysis of the atomic structure at and around the Type-90 RP planar defect. (A)** STEM-ADF image of a representative grain oriented in the $[1\bar{1}0]$ zone axis in Cs-1.1 film. **(B)** Atomic-resolution STEM-ADF image of the RP defect from the region marked in **(A)**. **(C)** Intensity line profile measured from the rectangular region highlighted in **(B)**. Intensity is integrated across the width of the rectangle to enhance signal. **(D)** Cs displacement vector map calculated from measurements of Cs-Cs column distance in **(B)** revealing shifts of Cs atoms. **(E)** Atomic model of γ-$CsPbI_3$ in the $[1\bar{1}0]$ zone axis. Yellow ovals indicate the projected shape of Pb/I columns in the tilted octahedra. **(F)** Ellipticity vector map measured from Pb/I columns in **(B)** revealing relaxation of octahedral tilt. Line direction indicates orientation of the ellipse major axis. Line length and colour indicates the magnitude of ellipticity (major axis/minor axis).

**(G)** Measured atomic model of the Type-90 RP planar defect showing the gap, Cs displacement, octahedral relaxation and [001]/[110] axis switch.

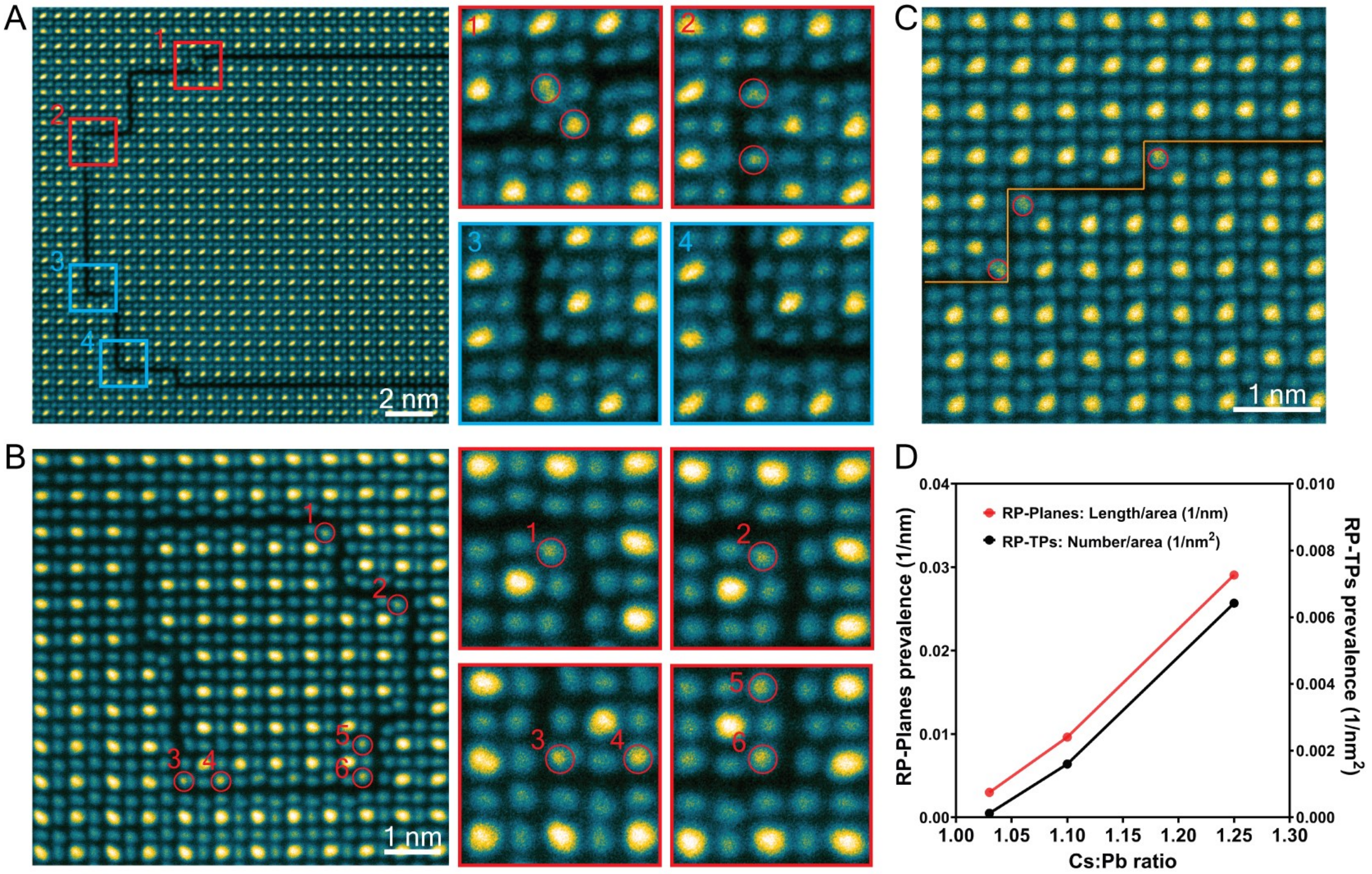

**Fig. 3. (A-C) Antisite defects at RP turning points in Cs-1.1 films;** STEM-ADF images showing the $Pb_{Cs}$ antisite defects at turning points highlighted in red (but absent in blue). **(A)** Type-0 RP defect. **(B)** Type-0 RP defect forming a loop. **(C)** Type-90 RP defect. **(D)** Prevalence of RP planes and RP turning points in different Cs:Pb ratio films.

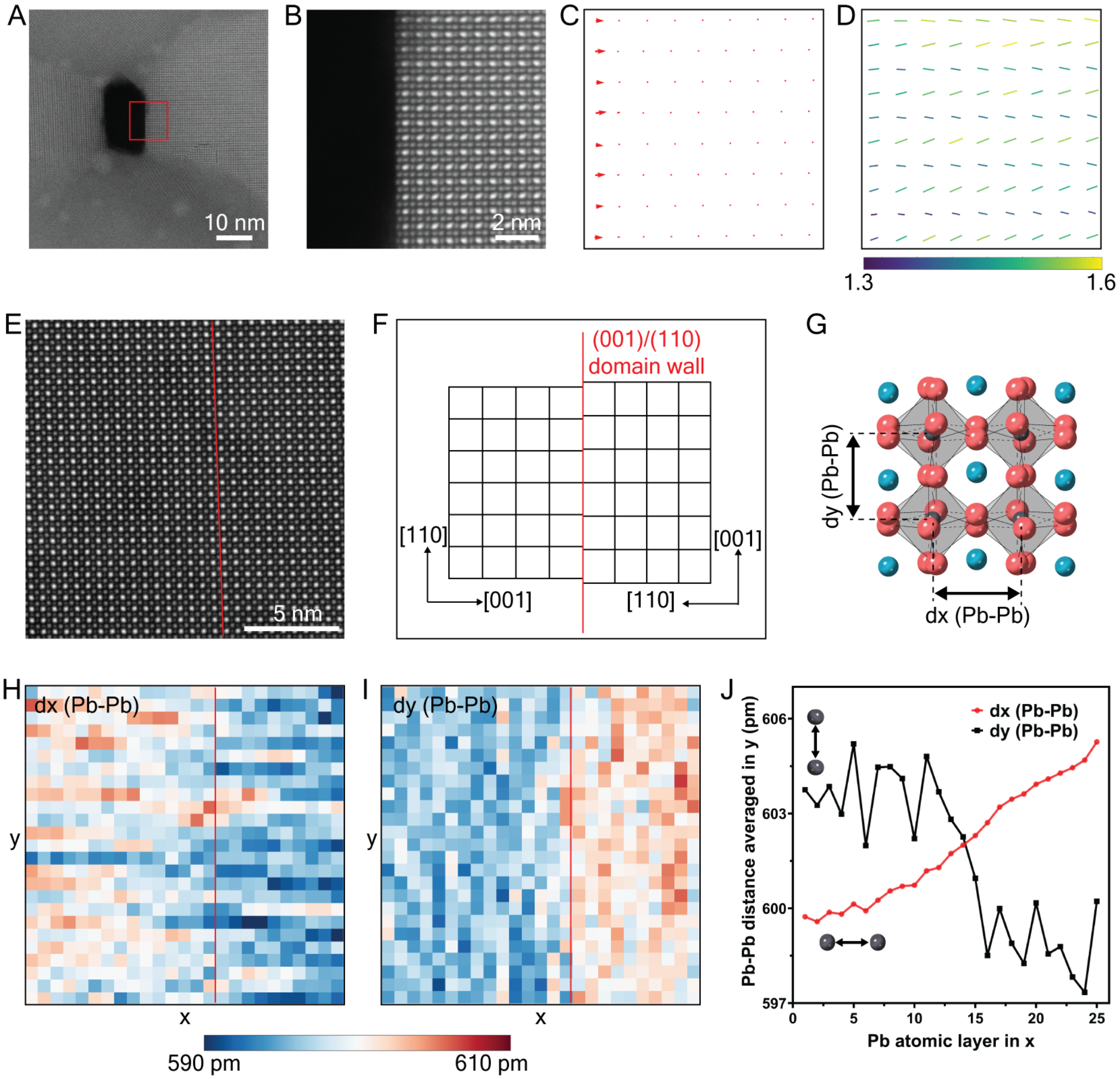

**Fig. 4. Insights into RP defect formation. Structure relaxation at a surface and strain at gap-less (001)/(110) boundary (A)** STEM-ADF image of a pinhole in the Cs-1.1 film in the $[1\bar{1}0]$ zone axis. **(B)** STEM-ADF image of the surface region from the red box marked in **(A)**. **(C)** Cs displacement vector map at the surface in **(B)**, comparable to the RP plane. **(D)** Map of the orientation and magnitude of the ellipse major axis measured from the intensity distribution at Pb/I columns in **(B)** revealing relaxation of octahedral tilt comparable to RP plane. Line direction indicates orientation of the ellipse major axis. Line length and colour indicates the magnitude of ellipticity (major axis / minor axis). **(E)** STEM-ADF image of $CsPbBr_3$ film oriented in the $[1\bar{1}0]$ zone axis showing a gap-less (001)/(110) domain boundary. **(F)** Schematic diagram illustrates the lattice mismatch at the domain boundary. **(G)** Lattice strain near the boundary can be measured from the distance of Pb/I columns (denoted as Pb-Pb distance) in directions parallel (defined as

the x direction) and perpendicular (defined as the y direction) to the boundary. **(H)** Pb-Pb distance measured in the x direction and **(I)** y direction. **(J)** Pb-Pb distance measurements in **(H, I)** averaged in the y direction.

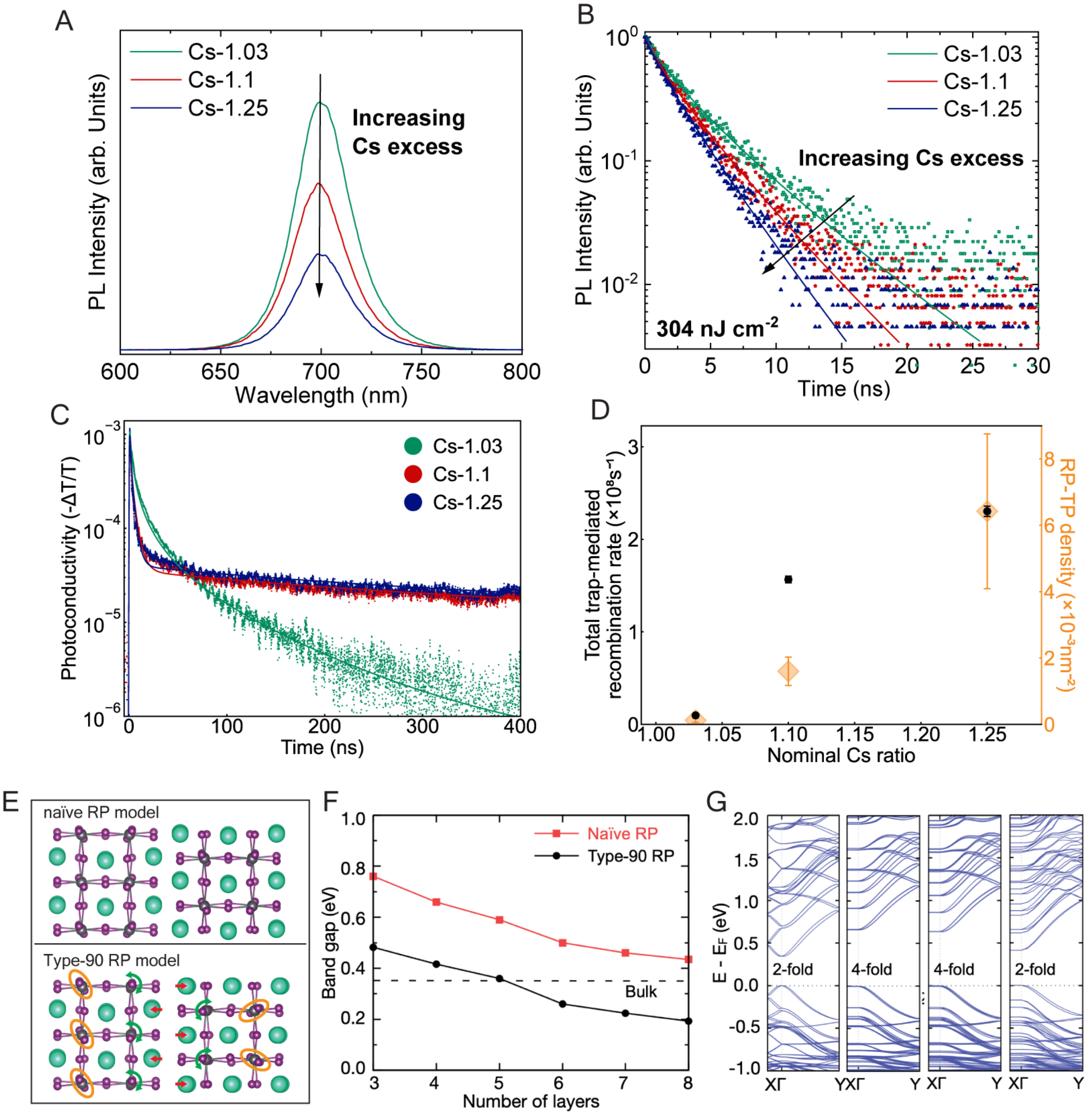

**Fig. 5. Photo-physical characterisation and density-function theory simulation of $CsPbI_3$.** (**A**) Photoluminescence (PL) spectra of 35 nm thick $CsPbI_3$ films of various nominal Cs:Pb ratios co-deposited on z-cut quartz substrate, following photo-excitation with a 398 nm-wavelength continuous wave laser; (**B**) Time-correlated single photon counting of the same films, after excitation with a 398 nm-wavelength pulsed diode laser at a repetition frequency of 10 MHz, and fits (solid lines) to transients with a stretched exponential model; (**C**) Photoconductivity transients measured via time-resolved microwave conductivity (TRMC) for films excited with a fluence of 32 μJcm$^{-2}$ (concurrently acquired time-resolved PL (TRPL) transients are shown in **Fig. S19**). Films were prepared as described above (**Fig. 5A**), and all photo-physical measurements were performed with samples held in $N_2$ atmosphere. TRMC and TRPL transients were simultaneously fitted with a dynamic recombination model, with the results plotted as solid lines; (**D)** The total trap-mediated recombination rate extracted by fitting the dynamic recombination model to concurrently acquired TRMC and TRPL transients, plotted alongside the prevalence of RP-turning points as measured via STEM-ADF, both as a function of the nominal Cs:Pb ratio; (**E)** RP planar defect atomic models; ‘real’ model of Type-90° defect using measured atomic positions incorporating octahedral relaxation, Cs displacement and 90°crystal rotation, and ‘naïve model’ without octahedral relaxation, Cs displacement (i.e. using same atomic positions as domain interior) nor 90°crystal rotation. **(F)** Calculated DFT band gaps for real models (black circles) and naïve models (red squares) as a function of the number of octahedral layers in the slab. The band gap of bulk orthorhombic $CsPbI_3$ is shown in dotted line for comparison. **(G)** Band structures calculated from DFT, including spin-orbit coupling, of a 4-layer defect interface model and an 8 × 2 × 2 $CsPbI_3$ supercell. Direction X to Γ in reciprocal space corresponds to the direction with the largest lattice parameter in real space (perpendicular to the RP defect), while Γ to Y corresponds to an in-plane direction in real-space. From left to right are results on: Bulk model, Naïve RP model, Type-0 RP model and Type-90 RP model.